\documentclass{ws-ijmpcs}
\usepackage{braket,psfrag}

\def\trimmarks{}

\allowdisplaybreaks

\newcommand{\VC}[1]{%
  \begin{tabular}[c]{l}%
    #1%
  \end{tabular}
}

\DeclareRobustCommand*\diff[2][]{%
   \mathop{
     \if@DIFF@roman 
        \mathrm{d}^{#1}
     \else 
        {d}^{#1}
     \fi
     \mskip-0.2\thinmuskip
   #2}\nolimits
}

\let\oldleft\left
\def\xleft{\mathopen{}\oldleft}

\newcommand{\trace}{\mathop{\mathrm{Tr}}}

\newcommand{\eqbad}{\stackrel{\textrm{?}}{=}}
\newcommand{\eqdef}{\stackrel{\textrm{def}}{=}}

\newcommand{\T}[1]{\boldsymbol{#1}_{\text{T}}}
\newcommand{\Tj}[2]{\boldsymbol{#1}_{#2\,\text{T}}}

\def\Tsub#1_#2{\Tj{#1}{#2}}
\newcommand{\Tsc}[1]{#1_{\text{T}}}

\begin{document}

\markboth{Collins}{New definition of TMD parton densities}

\catchline{}{}{}{}{}
%
%%%%%%%%%%%%%%%%%%%%%%%%%%%%%%%%%%%%%%%%%%%%%%%%%%%%%%%%%%%%%%%%%%%%

\title{NEW DEFINITION OF TMD PARTON DENSITIES}

\author{JOHN COLLINS}
\address{Physics Department, Penn State University, 
    University Park PA 16802, U.S.A.
\\
   collins@phys.psu.edu
}

\maketitle

%====================================
\begin{history}
%\received{Day Month Year}
%\revised{Day Month Year}
\end{history}

%====================================
\begin{abstract}
%\keywords{Keyword1; keyword2; keyword3.}
I give an account of a new definition of transverse-momentum-dependent
parton densities.  The new definition solves a number of difficulties
and inconsistencies in earlier definitions.
\end{abstract}

%\ccode{PACS numbers: 11.25.Hf, 123.1K}

%====================================
\section{Introduction}

In this talk, I presented a new definition of
transverse-momentum-dependent (TMD) parton densities that I developed
in my recent book.\cite{Collins:2011qcdbook} They overcome many
deficiencies of previous definitions.  Numerical values for these
densities with the new definition were obtained by Aybat and
Rogers\cite{Aybat:2011zv} from fits to data.

Before embarking on the QCD definition, I give a reminder about how
TMD parton densities arise naturally in the parton model.  Then I
summarize some of the difficulties that have to be overcome in
defining TMD parton densities in QCD.  After that I present the new
definition, compare them with other work, and indicate some of the
implications of the new definition.

A number of assertions are given here without justification.  For
more details on the justification, see Ref.\
\refcite{Collins:2011qcdbook}.

%====================================
\section{Basics of TMD parton densities}

\begin{figure}
\centering
  \includegraphics[scale=0.4]{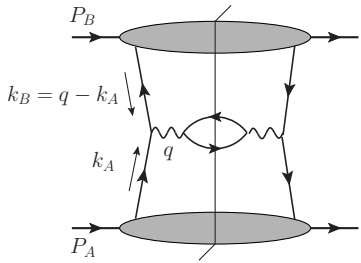}
\caption{The parton model for the Drell-Yan process
\label{fig:DY.pm}}
\end{figure}

A motivation that TMD parton densities are useful objects to use is
provided by the parton model for the Drell-Yan process, i.e., the
production of a high-mass lepton pair in a high-energy hadron-hadron
collision.  In the parton model, the lepton pair is formed by
annihilation of a quark in one hadron and an antiquark from the other,
Fig.\ \ref{fig:DY.pm}.  This gives TMD factorization:
\begin{equation}
\label{eq:DY.pm.fact}
  \frac{ \diff{\sigma} }{ \diff[4]{q}\diff{\Omega} }
  ~\eqbad~ \sum_j \int \diff[2]{\Tj{k}{A}} 
       ~f_{j/h_A}(x_A,\Tj{k}{A})
       ~f_{\bar{j}/h_B}(x_B,\T{q}-\Tj{k}{A})
       ~\frac{ \diff{\hat{\sigma}_{j\bar{j}}} }{ \diff{\Omega} },
\end{equation}
where the $f_{j/H}$ factors are the TMD parton densities, functions of
a longitudinal momentum fraction and a transverse momentum, while
$\diff{\hat{\sigma}}$ is the partonic hard scattering in the lowest-order
approximation.  The query over the equality sign is a reminder that
this is not exactly the correct TMD factorization theorem in QCD.  

Because the transverse momentum $\T{q}$ of the lepton pair is the sum
of the transverse momenta of the partons, the cross section
\eqref{eq:DY.pm.fact} is directly sensitive to partonic transverse
momenta.  If the partons had no transverse momentum, the cross section
would be a delta-function at $\T{q}=0$.  This contrasts with
deep-inelastic scattering where only one parton participates in the
hard scattering, so that its transverse momentum can be neglected with
respect to the large momentum transfer $Q$ in the hard scattering.

The need for TMD parton densities in Eq.\ (\ref{eq:DY.pm.fact})
establishes that TMD parton densities are important quantities for
a quantitative description of many hard processes.

To derive the parton model, one needs to use a cancellation of
spectator-spectator interactions.\cite{Cardy:1974vq}  In addition one
needs to assume other topologies of graph are unimportant, that
partonic $\Tsc{k}$ and virtuality are limited, and that no
higher-order corrections are needed to the hard scattering.  All of
the last three assumptions are violated in QCD and are associated with
a need to modify the definitions of the parton densities and the
factorization formula in QCD.

%-------------------------------------------
\subsection{Explicit definition of TMD parton density: complications
  in QCD} 

In constructing an operator definition of a TMD parton density in a
hadron, I assume that the hadron is moving in the $+z$ direction, and
I will use light-front coordinates defined by $v^\mu=(v^+,v^-,\T{v})$,
with $v^{\pm}=(v^0\pm v^z)/ \sqrt2$, $\T{v}=(v^x,v^y)$.

\begin{figure}
  \centering
  $\lim\limits_{L\to\infty}\VC{\includegraphics[scale=0.8]{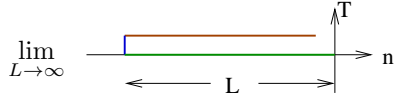}}$
  \caption{Wilson line in naive definition of gauge invariant parton
    density, viewed from the side. } 
  \label{fig:TMD.pdf.WL}
\end{figure}

The parton model leads to a definition of a parton density as a hadron
expectation value of the number density of a parton, as specified in
light-front quantization.  A first attempt at applying this in QCD
uses the $A^+=0$ gauge.  This is equivalent to the following
gauge-invariant definition with a Wilson line in the
direction\footnote{Note that the derivation of factorization requires
  that parton densities for the Drell-Yan process use past-pointing
  Wilson lines.\cite{Collins:2004nx}}  $-n=-(0,1,\T{0})$:
\begin{multline}
\label{eq:TMD.pdf.pm.def}
  f_{j/h}(\xi,\T{k})
  \eqbad
     \int \frac{ \diff{k^-} }{ (2\pi)^4 }
           \trace \frac{\gamma^+}{2} 
       \hspace*{3mm}
       \VC{\includegraphics[scale=0.45]{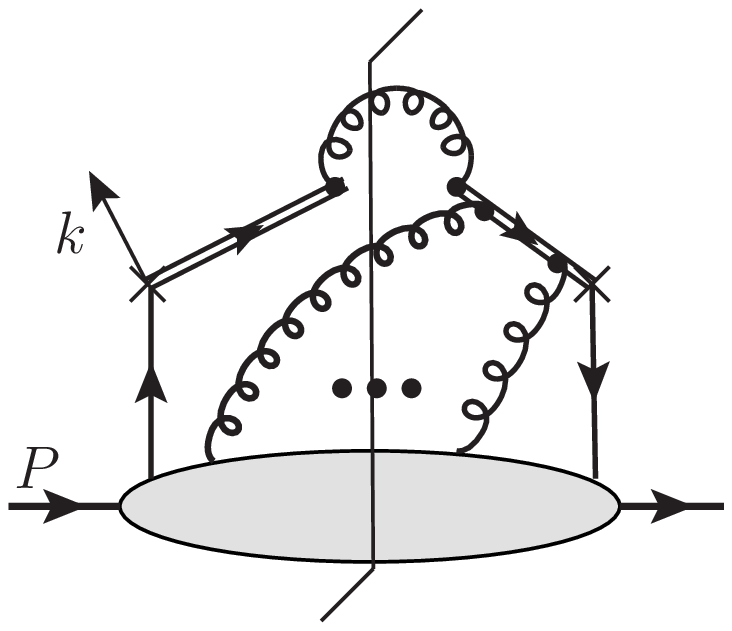}}
\\
=
  \text{F.T.}
    \braket{P| \,
            \overline{\psi}_j(0,w^-,\T{w}) \,
            W[w, -n]^\dag\,
            \frac{\gamma^+}{2} \,
            [\mbox{tr.\ link}] \,
            W[0, -n]          
       \psi_j(0) \,
    |P}_{\text{c}} ,
\end{multline}
where the Wilson line is
\begin{equation}
\label{eq:wildef}
W(\infty ,x;-n) = P \exp \left[- ig_0 \int_0^\infty d s \; (-n) \cdot A_0^a (x - s n) t^a \right],
\end{equation}
``F.T.'' denotes a Fourier transform with respect to $w^-$ and
$\T{w}$, and ``$[\mbox{tr.\ link}]$'' denotes a transverse Wilson line
at infinity.  The product of Wilson lines is a path-ordered
exponential of gluon fields along the line shown in Fig.\
\ref{fig:TMD.pdf.WL}.

There are several complications that prevent this definition from
being taken literally:
\begin{itemize}
\item UV divergences.  These are removed by suitable renormalization
  counterterms.
\item Rapidity divergences.\cite{Collins:2003fm} These concern gluons
  whose rapidity goes to $-\infty$.
\item There are Wilson line self energies.  These are present in the
  above definition of the TMD parton density, but not in the actual
  cross section.  They create divergences as the length of the Wilson
  line goes to infinity.\cite{Bacchetta:2008xw}
\end{itemize}

%-------------------------------------------
\subsection{Example: Regions for one gluon}

\begin{figure}
  \centering
  \begin{tabular}{cccc}
  \VC{\includegraphics[scale=0.4]{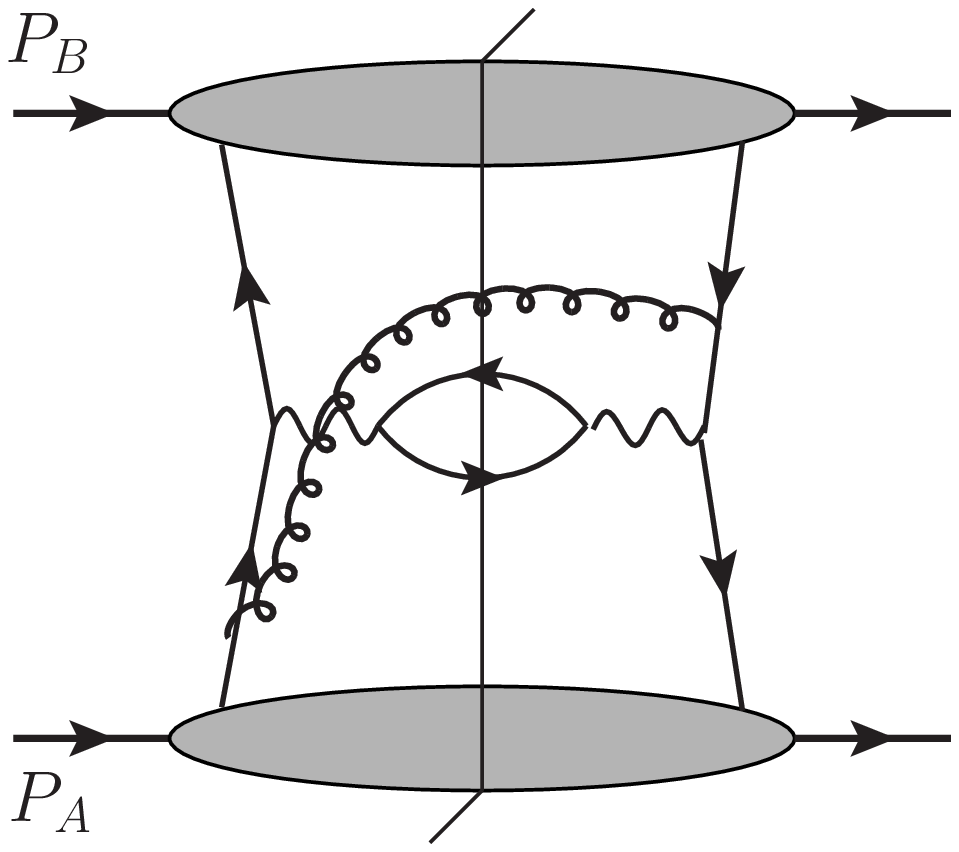}}
  &
  \raisebox{-5mm}{\VC{\includegraphics[scale=0.45]{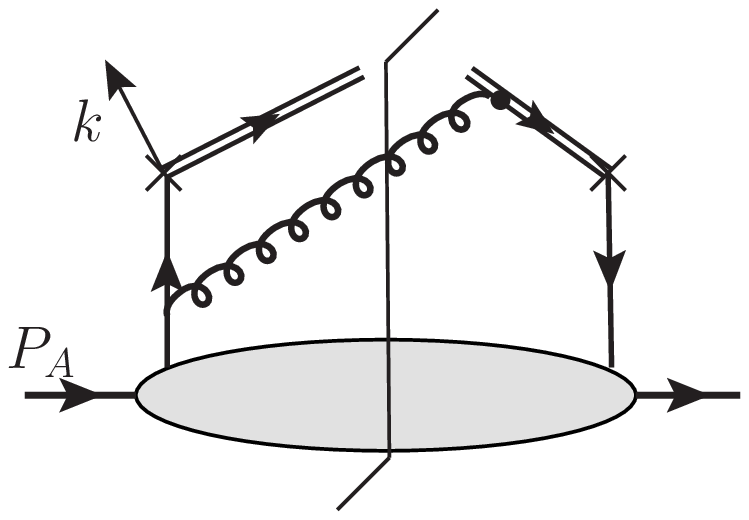}}}
  \\
  (a) & (b)
  \end{tabular}
  \caption{Graphs for the Drell-Yan cross section and for a
    TMD parton density in hadron $A$.}
  \label{fig:DY.graphs}
\end{figure}

These complications are illustrated by the graph in Fig.\
\ref{fig:DY.graphs}(a), where into a parton-model graph for the
Drell-Yan process is inserted one gluon exchanged between the
annihilating quark and antiquark.  Associated with this in the
factorization of collinear configurations is Fig.\
\ref{fig:DY.graphs}(b) for the parton density in hadron $A$ and a
similar graph for the parton density in hadron $B$.

\begin{figure}
  \centering
  \psfrag{y}{\small $y$}
  \psfrag{yA}{\small $y_{p_A}$}
  \psfrag{yB}{\small $y_{p_B}$}
  \psfrag{ln kT/m}{\small $\ln(\Tsc{k}/m)$}
  \psfrag{k-~pB-}{\small \hspace*{-15mm}$k^-\sim p_B^-$}
  \psfrag{k+~pA+}{\small $k^+\sim p_A^+$}
  \psfrag{k-~pB-}{}
  \psfrag{k+~pA+}{}
  \setlength{\tabcolsep}{5mm}
  \begin{tabular}{cccc}
  \includegraphics[scale=0.6]{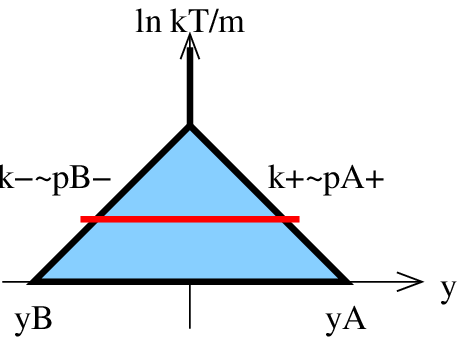}
  &\includegraphics[scale=0.6]{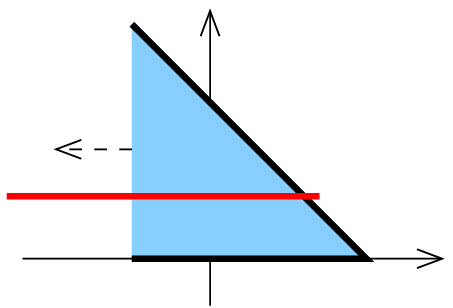}
  &\includegraphics[scale=0.6]{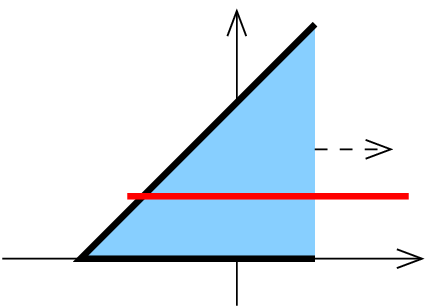}
  \\
  (a) & (b) & (c)
\end{tabular}
\caption{Regions of gluon momentum dominating in the graphs of Fig.\
  \ref{fig:DY.graphs} and in the corresponding graph for the TMD
  parton density in hadron $B$.}
  \label{fig:DY.regions}
\end{figure}

These graphs give leading power contributions from regions of gluon
momentum indicated by the shaded/blue regions in Fig.\
\ref{fig:DY.regions}, a plot of the space of gluon rapidity and the
logarithm of its transverse momentum.  A frame is used in which hadron
$A$ moves in the $+z$ direction and hadron $B$ in the $-z$ direction.
The red line indicates roughly the region needed when the transverse
momentum $\Tsc{q}$ of the lepton pair is fixed.  The leading logarithm
approximation is obtained by replacing the integrand in
$\int\diff{\ln\Tsc{k}}\diff{y} \ldots$ by a constant.  In the cross section
integrated over $\Tsc{q}$, the triangular region in the first plot
gives the well known Sudakov double logarithm of $Q$ for Fig.\
\ref{fig:DY.graphs}(a).  The edges of the triangle are set roughly by
the rapidities of the annihilating quark and antiquark, and by an
infra-red cutoff on gluon $\Tsc{k}$.  In massless perturbation theory,
with $m_g=0$, an extra doubly logarithmic contribution comes from even
smaller $\Tsc{k}$.

Replacing the gluon and the lower part of the graph by a corresponding
graph for the parton density in hadron $A$, Fig.\
\ref{fig:DY.graphs}(b), gives the region shown in Fig.\
\ref{fig:DY.regions}(b).  The parton density graph corresponds to a
good approximation for the gluon in Fig.\ \ref{fig:DY.graphs}(a) when
it is collinear to hadron $A$, i.e., when $y$ is sufficiently positive
and $\Tsc{k}$ is sufficiently much less than $Q$.  The proof of this
assertion uses a Ward identity to move the upper end of the gluon line
in Fig.\ \ref{fig:DY.graphs}(a) onto a Wilson line.

But the approximation becomes very bad for negative $y$, so much so,
that there is a divergence in the integral to $y=-\infty$ (a ``rapidity
divergence'').  

Similarly, in Fig.\ \ref{fig:DY.regions}(c), the corresponding gluonic
term in the parton density in hadron $B$ reproduces the region of
gluon momentum collinear to hadron $B$, but fails badly in the
positive rapidity region.

In the QCD factorization theorem that generalizes Eq.\
\eqref{eq:DY.pm.fact}, these two parton-density contributions are to
be added.  A further problem is then that the region of central gluon
rapidity is double counted.

All of the problems are overcome by finding a suitably modified
definition of the TMD parton densities, together with a valid
factorization theorem using the parton densities.

%====================================
\section{TMD factorization in QCD}

TMD factorization in QCD has the general form given by Collins, Soper,
and Sterman\cite{Collins:1984kg} (CSS):
\begin{equation}
\label{eq:CSS.fact}
    \diff{\sigma} = H \times \mbox{convolution of } A B S
           ~+~ \mbox{high-$\Tsc{q}$ correction ($Y$)}
           ~+~ \mbox{power-suppressed}.
\end{equation}
Here, $H$ is a hard-scattering factor, $A$ and $B$ are TMD
parton densities, and $S$ is a soft factor, with the product $ABS$
being a convolution in transverse momentum.  Double counting
subtractions are used in the definition of the factors, and there must
be present some kind of cut off on rapidity divergences. 

The $HABS$ part correctly gives the cross section when $\Tsc{q}\ll Q$,
but it fails at larger $\Tsc{q}$.  The $Y$ term corrects the errors at
large $\Tsc{q}$; it has the form given by ordinary collinear
factorization, but with a subtraction of the low $\Tsc{q}$ region.

All errors in Eq.\ \eqref{eq:CSS.fact} are suppressed by a power of
$1/Q$.  The pattern of errors in the individual terms is interesting:
\begin{equation}
    \text{Errors in } 
    \left\{ \begin{tabular}{c}
              collinear factorization \\ 
              TMD factorization ($HABS$) \\
              TMD factorization $+~Y$ 
           \end{tabular}
    \right\} 
    \text{ are a power of } 
    \left\{ \begin{array}{c}
               \dfrac{\Lambda}{\Tsc{q}};\\
               \dfrac{\Lambda}{Q},~\dfrac{\Tsc{q}}{Q};\\
               \dfrac{\Lambda}{Q}.
           \end{array}
    \right. 
\end{equation}

Each of the $A$, $B$, and $S$ factors has a non-perturbative
contribution that cannot currently be predicted from the theory; to
determine the non-perturbative contributions one must fit them to
data.  Unfortunately, the soft factor $S$ always appears multiplied by
two collinear factors (not only in the Drell-Yan process, but also in
similar factorization theorems for other reactions).  Thus there is no
possibility of measuring $S$ independently of the parton density
factors $A$ and $B$.

A better formulation can therefore be obtained by redefining the
parton density factors so that they each incorporate a square root of
the soft factor.  The new definition, presented below, accomplish
this, albeit somewhat indirectly.

%====================================
\section{New definition}

The new definition is in terms of unsubtracted quantities that have
non-light-like Wilson lines.  The unsubtracted parton density in
hadron $A$ has a Wilson line of rapidity $y_2$:
\begin{equation}
\label{eq:TMD.pdf.unsub}
  \tilde{f}_{f/H_A}^{\text{unsub}}(x,\T{b};y_{P_A}-y_2)
  \eqdef{}
  \trace_{\text{color}} \trace_{\text{Dirac}} \frac{\gamma^+}{2}
   \int \frac{ \diff{k^-} \diff[2-2\epsilon]{\T{k}} }{ (2\pi)^{4-2\epsilon} }
  ~ 
  e^{-i \T{k} \cdot \T{b} }
  \VC{\includegraphics[scale=0.4]{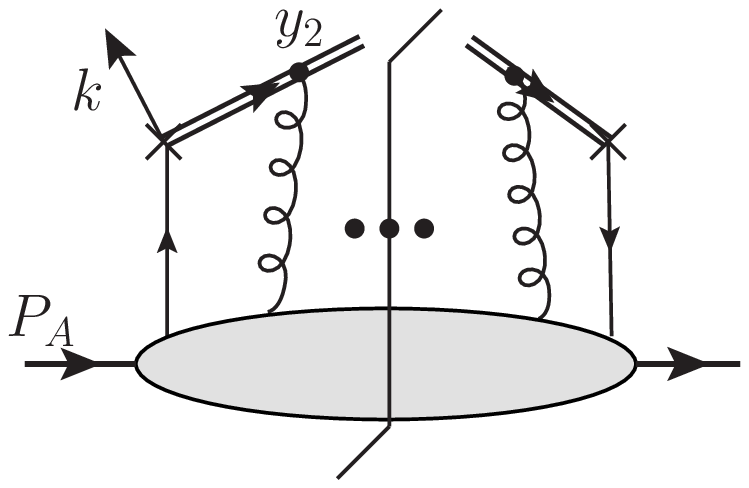}}
\end{equation}
The unsubtracted soft factor has Wilson lines of rapidities $y_1$ and
$y_2$, with $y_1>y_2$:
\begin{equation}
\label{eq:S.unsub}
   \tilde{S}(\T{b}) \eqdef{}
    \frac{1}{N_c}
    \int \frac{ \diff[4-2\epsilon]{k_S} }{ (2\pi)^{4-2\epsilon} }
    e^{-i\Tj{k}{S}\cdot\T{b}}
    \quad 
       \VC{\includegraphics[scale=0.3]{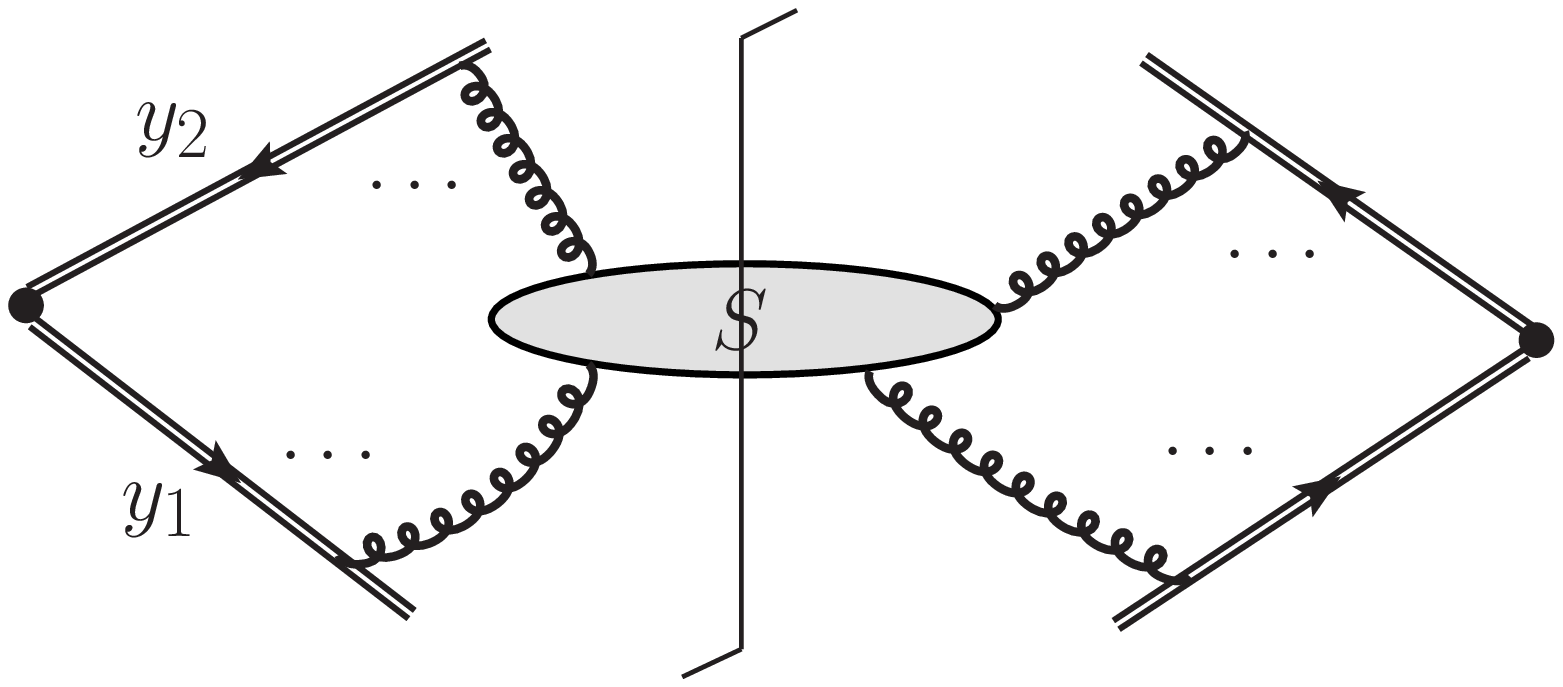}}
\end{equation}
The definition of the TMD parton density uses an auxiliary rapidity
parameter $y_n$ which sets the finite rapidity of certain Wilson
lines.  The rapidities of other Wilson lines are taken to $+\infty$ or
$-\infty$: 
\begin{multline}
\label{eq:TMD.pdf.defn}
    \tilde{f}_{f/H_A}(x,\T{b};\zeta_A;\mu) 
    \eqdef
    \lim_{\substack{y_1\to+\infty \\ y_2\to-\infty}}
    \tilde{f}_{f/H_A}^{\text{unsub}}(x,\T{b};y_{P_A}-y_2) 
    \sqrt{
         \frac{ \tilde{S}(\Tsc{b},y_1,y_n) }
              { \tilde{S}(\Tsc{b},y_1, y_2) ~ \tilde{S}_{(0)}(\Tsc{b},y_n, y_2) }
    }
\\
    =
    \tilde{f}_{f/H_A}^{\text{unsub}}\big( x,\T{b};y_{p_A}-(-\infty) \bigr) 
    \sqrt{
         \frac{ \tilde{S}(\Tsc{b};+\infty,y_n) }
              { \tilde{S}(\Tsc{b};+\infty,-\infty) ~ \tilde{S}_{(0)}(\Tsc{b};y_n,-\infty)}
    }.
\end{multline}
Here $\zeta_A$ is defined to be $M_{P_A}^2x^2e^{2(y_{P_A}-y_n)}$ to match
a definition in Ref.\ \refcite{Collins:1981uw}.  A UV renormalization
factor is implicit, as is the limit $\epsilon\to0$ for the removal of the
dimensional regularization of UV divergences.

The multiplications, divisions, and square roots are applied to the
various factors in transverse coordinate space.  They would correspond
to complicated convolutions were they to be applied in transverse
momentum space.

Diagrammatically, this definition can be written as
\begin{equation}
  \VC{\includegraphics[scale=0.4]{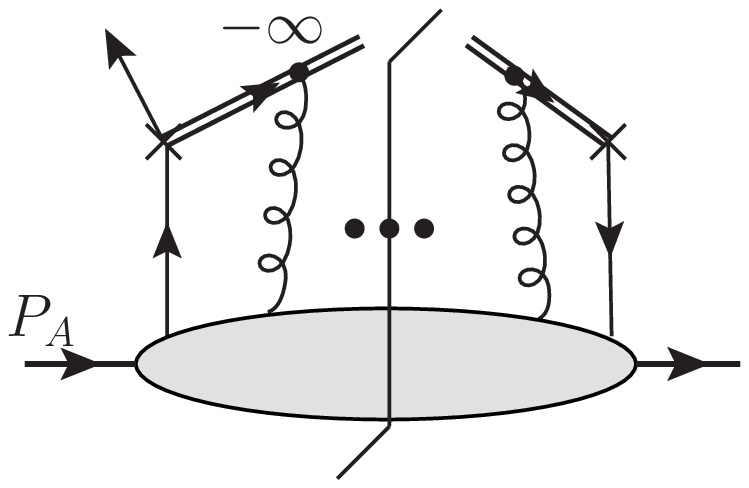}}  
  \hspace*{4mm}
  \sqrt{ \frac{
          \VC{\includegraphics[scale=0.2]{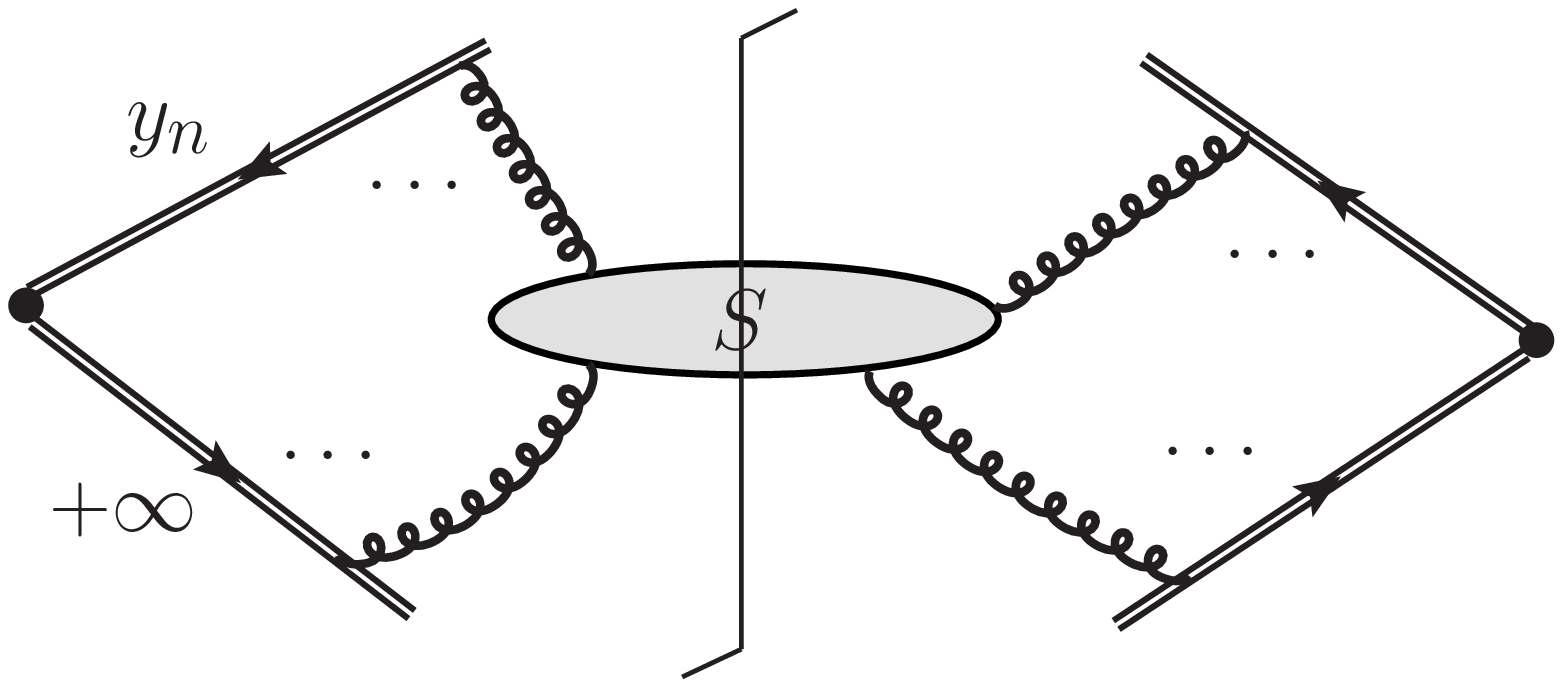}}
       }
       {
          \VC{\includegraphics[scale=0.2]{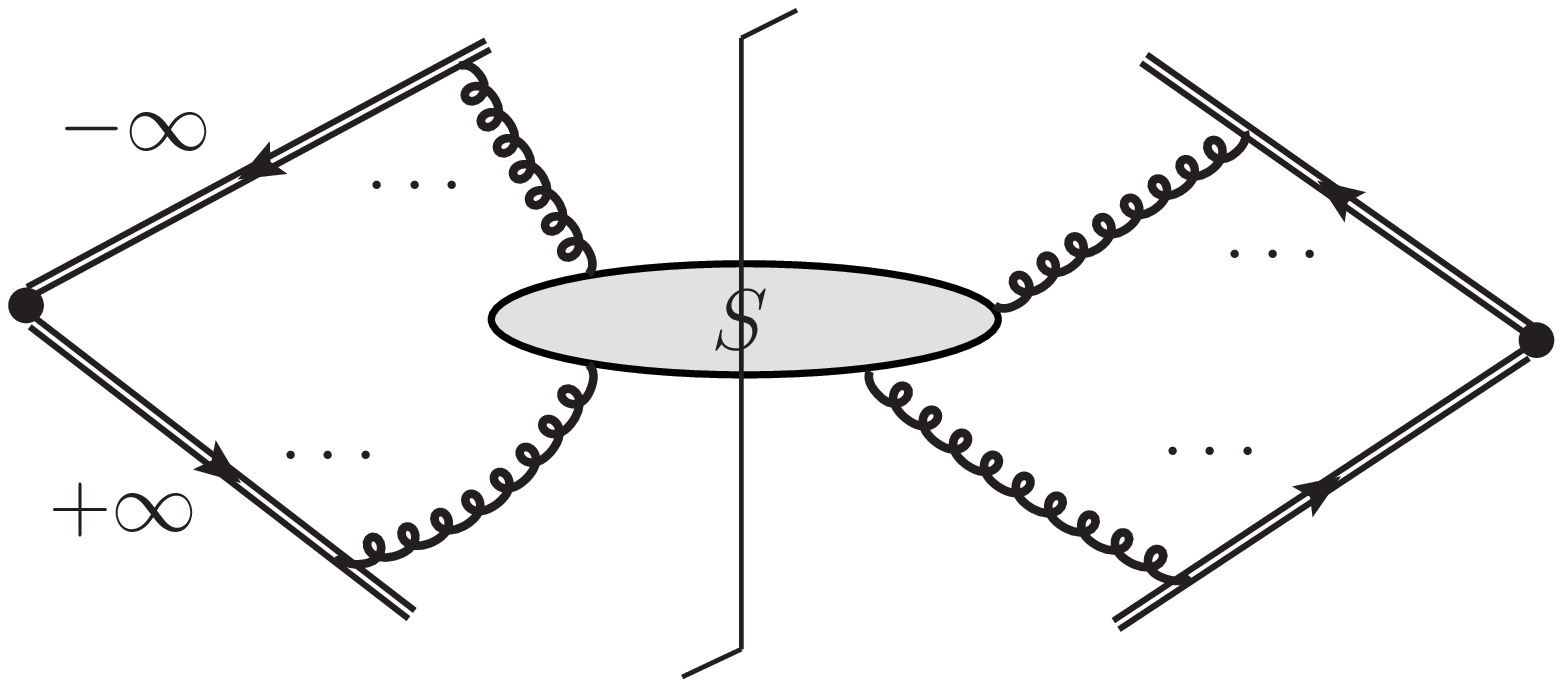}}
          \VC{\includegraphics[scale=0.2]{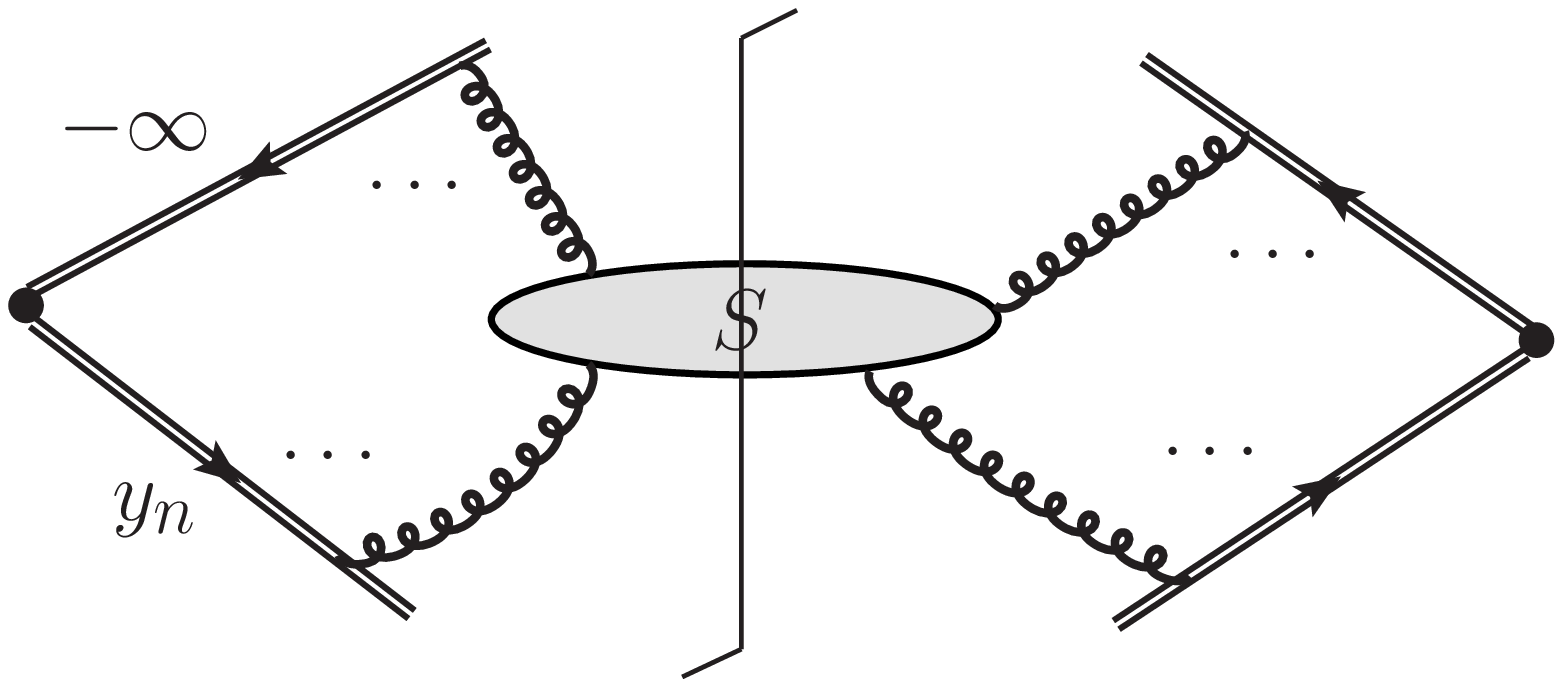}}
       }
  }
\end{equation}

%-------------------------------------------
\subsection{Why this strange definition?}

This definition seems unexpectedly complicated, although the square
root is to be expected given the motivation of absorbing a square root
of the soft factor into each parton density in the factorization
property.  In fact, the definition is unique (up to the choice of the
prescription for renormalization of UV divergences) given the
following properties.
\begin{itemize}
\item The parton density (in transverse coordinate space) is the
  product of the basic parton density and powers of the soft factor.
\item Light-like Wilson lines are used if possible:
  \begin{itemize}
  \item The basic parton density has only a light-like Wilson line
  \item Each soft factor has at most one non-light-like line
  \end{itemize}
\item Rapidity divergences cancel, as do divergences when the length
  of the Wilson lines goes to infinity ($L\to\infty$).
\end{itemize}
Note that only factorization-compatible definitions should be
considered, that the above definition is factorization-compatible, and
that the definition implements double-counting subtractions between
the soft region and the two collinear regions of momentum space.

%-------------------------------------------
\subsection{Consequences of the strange definition}

Some consequences of the new definition are:
\begin{itemize}

\item Factorization holds, as in Eq.\ \eqref{eq:CSS.fact}, but without
  soft factor.

\item Thus the TMD part of factorization is just like that in the
  parton model, \eqref{eq:DY.pm.fact}, except that the parton
  densities depend on $\zeta_A$ (and a corresponding variable $\zeta_B$ for
  the parton density in hadron $B$) and on $\mu$, and except that the
  hard scattering has higher order corrections.  Note that $\zeta_A\zeta_B\simeq
  Q^4$.

\item Rapidity divergences and Wilson-line self-energy divergences
  cancel. 

\item Contributions from the gauge link at infinity cancel in Feynman
  gauge. 

\item There is an effective cutoff on gluon rapidity at $y_n$

\item When the energy of an experiment and the dilepton mass are
  scaled up, the parton densities change because $\zeta_A$ and $\zeta_B$
  increase.  The dependence on these variables is governed by the
  dependence on $y_n$, the rapidity of the non-light-like Wilson
  lines, and this dependence is governed by the CSS evolution
  equations to be discussed below.

  In addition, the renormalization scale $\mu$ should be of order $Q$
  to avoid large logarithmic higher-order corrections in the hard
  scattering.  The $\mu$-dependence is governed by a
  renormalization-group equation.

\item The versions of the CSS evolution equations for the newly
  defined parton densities are simpler than the original
  ones.\cite{Collins:1984kg} In particular, they no longer have
  hard-to-control power-law corrections.

\end{itemize}

%====================================
\section{Evolution, etc}

In this section, I summarize the evolution equations and the
small-$\Tsc{b}$ expansions.  These enable the TMD parton densities to
be expressed in terms of (a) non-perturbative quantities without scale
dependence, and (b) perturbative quantities that do not have large
logarithms in the coefficients of their perturbative expansions. 

The dependence of the parton density \eqref{eq:TMD.pdf.defn} on $\zeta_A$
is obtained by differentiating with respect to $y_n$ the two soft
factors that use a Wilson line of rapidity $y_n$.  This gives a
CSS-style equation
\begin{equation}
\label{eq:CSS.evol}
  \frac{ \partial \ln \tilde{f}_{f/H_A}(x,\Tsc{b}; \zeta; \mu) }
       { \partial \ln \sqrt{\zeta} }
  = 
  \tilde{K}(\Tsc{b};\mu).
\end{equation}
The renormalization group (RG) equation of the kernel of this equation is
\begin{equation}
\label{eq:K.RG}
  \frac{ \diff{\tilde{K}} }{ \diff{\ln \mu } }
  = -\gamma_K\xleft(g(\mu)\right).
\end{equation}
The RG equation of the TMD parton density is
\begin{equation}
\label{eq:TMD.pdf.RGE}
  \frac{ \diff{ \ln \tilde{f}_{f/H}(x,\Tsc{b};\zeta;\mu) }}
       { \diff{\ln \mu} }
  = \gamma\big( g(\mu);\zeta/\mu^2 \bigr).
\end{equation}
From these equations, follows an equation for the $\zeta$-dependence of
$\gamma$, and hence that
\begin{equation}
    \gamma\big( g(\mu); \zeta/\mu^2 \bigr)
    = \gamma( g(\mu); 1 )
      - \frac12 \gamma_K(g(\mu)) \ln \frac{ \zeta }{ \mu^2 }.
\end{equation}
At small $\Tsc{b}$, the TMD parton density has a factorized formula in
terms of ordinary integrated parton densities:
\begin{displaymath}
  \tilde{f}_{f/H}(x,\Tsc{b};\zeta;\mu) 
  = \sum_j \int_{x-}^{1+} \frac{ \diff{\hat{x}} }{ \hat{x} }
       \,\tilde{C}_{f/j}\xleft( x/\hat{x},\Tsc{b};\zeta,\mu,g(\mu) \right)
       \, f_{j/H}(\hat{x};\mu)
~+~ O\xleft[(m\Tsc{b})^p \right]
\end{displaymath}
See Aybat's talk at this conference for results at level of TMD parton
densities.

%====================================
\section{Comparisons}

In this section, I compare the new definition with some other
definitions that have appeared in the literature.

%-------------------------------------------
\subsection{CSS-style definitions}

The new definition can be considered as a development of that
originally proposed by Collins and Soper.\cite{Collins:1981uw} The old
definition used the basic operator definition
\eqref{eq:TMD.pdf.pm.def}, but without a Wilson line and with the use
of a non-light-like axial gauge $n\cdot A=0$ to cut off rapidity
divergences.

With this definition there was a lack of an actual proof of TMD
factorization for the Drell-Yan process.\cite{Collins:1984kg} The
problem is that Glauber region is difficult to treat in ``physical
gauges'' such as the $n\cdot A = 0$ gauge. This lack has been remedied in
Ref.\ \refcite{Collins:2011qcdbook} with the aid of the new
definition.  The old definition can be converted to a gauge-invariant
form, with the use of non-light-like Wilson lines, which should be
space-like to get factorization.\cite{Collins:2004nx} However there
are divergences associated with self energies on the infinitely long
dipolar Wilson lines.  This last problem was only recently noticed, by
Bacchetta, Boer, Diehl, and Mulders --- see App.\ A of Ref.\
\refcite{Bacchetta:2008xw}.

Furthermore, with the Collins-Soper definition there was a separate
soft factor in the factorization formula, but as stated earlier, the
non-perturbative part of the soft factor cannot be independently
determined from data.  

Finally the derivation of the version of the evolution equation
\eqref{eq:CSS.evol} appropriate to the old definition involved
ignoring errors that are power-suppressed only at small transverse
momentum.  Since the TMD densities are also used at large transverse
momentum (in conjunction with the $Y$ term in \eqref{eq:CSS.fact}), it
follows that the actually-used TMD pdfs in the Collins-Soper method
are not actually those that Collins and Soper defined explicitly.

Ji, Ma, and Yuan\cite{Ji:2004wu} improved the Collins-Soper
definition of the TMD densities and the soft factor with the use of
multiple non-light-like Wilson lines.  However, they did not see how
to take infinite-rapidity limits as in Eq.\ \eqref{eq:TMD.pdf.defn},
which would have simplified their formalism.  Their factorization
formula still contains a soft factor.

%-------------------------------------------
\subsection{Naive light-cone-gauge definition}

A number of authors have defined a TMD density as the naive
expectation value of a light-front parton number operator in
light-cone gauge, i.e., by Eq.\ \eqref{eq:TMD.pdf.pm.def} in $A^+=0$
gauge with omission of the Wilson line.  Such a definition suffers
from rapidity divergences.\cite{Collins:2003fm} It can only be
considered a valid definition if a cutoff is applied.  But, as far as
I can see, the need for a cutoff is often not recognized, or, if it
is, a cutoff is applied in such an implicit way that it is difficult
to work out what the actual definition is.

Similar problems afflict much work that uses TMD densities at small
$x$ with BFKL-related approaches --- see Avsar's talk for more
details.

%-------------------------------------------
\subsection{Cherednikov and Stefanis's work}

In several recent papers, Cherednikov and
Stefanis\cite{Cherednikov:2007tw,Cherednikov:2008ua,Cherednikov:2009wk}
have presented an analysis of possible definitions of TMD parton
densities.  They have drawn conclusions about the anomalous dimensions
of the TMD parton densities and about the role of gauge links at
infinity.  Particular results concern the presence or absence of
rapidity divergences in certain versions of light-cone-gauge
quantization of QCD.  Their results are based primarily on one-loop
calculations.  In particle, they claim that the Mandelstam-Leibbrandt
prescription\cite{Mandelstam:1982cb,Leibbrandt:1983pj} is particularly
suitable, by giving finite results in situations where other methods
of using the light-cone gauge give divergences.

Their gauge-invariant definition is (slightly modified from)
\begin{equation}
  \tilde{f}_{f/H_A}^{\text{Cher.-Stef.}}(x,\T{b})
  = 
  \VC{\includegraphics[scale=0.4]{figures/pdf_quark_GI_tmd-A-i2_NO_k}}  
  ~\times~
   \left[ \VC{\includegraphics[scale=0.25]{figures/2PI-S-i1i2}}
   \right]^*
\end{equation}
with a certain treatment of gauge links at infinity.  In calculations
light-cone gauge is used, with a particular cutoff in the gluon
propagator:
\begin{equation}
\label{eq:ML}
     D^{\mu\nu}_{\text{ML}}
     =
     \frac{i}{q^2} 
     \left[ -g^{\mu\nu} + 
             (q^\mu n^\nu + n^\mu q^\nu)
             \frac12 \left( \frac{1}{ q^+ +i\eta } + \frac{1}{ q^+ -i\eta } \right)
     \right],
\end{equation}
where $\eta$ nonzero and positive.  This definition in fact has
uncanceled rapidity divergences from graphs that the authors appear
not to calculate.  Examples are
\begin{equation}
       \VC{\includegraphics[scale=0.4]{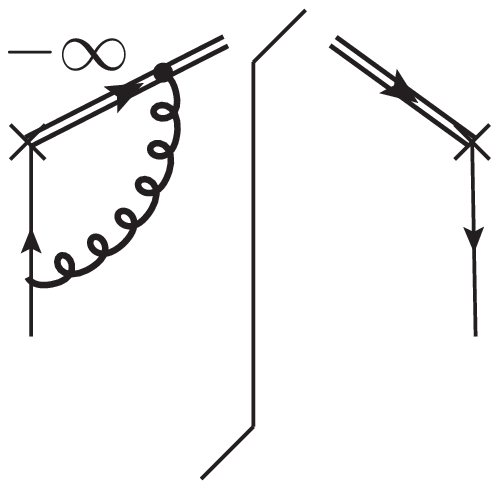}}
  \quad, \qquad
       \VC{\includegraphics[scale=0.4]{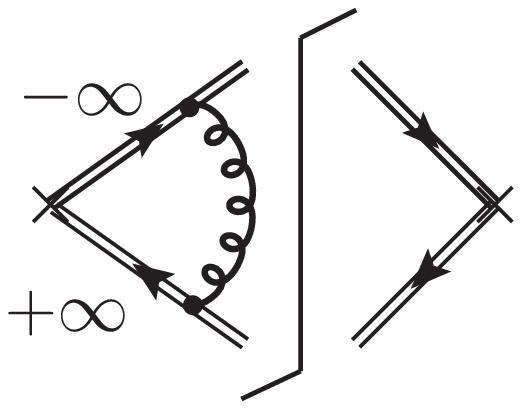}}  
\end{equation}
(Note that the regulated propagator \eqref{eq:ML} does \emph{not} obey
$n_\mu D^{\mu\nu}_{\text{ML}}=0$, so in these graphs the gluon attachments
to the Wilson line of negative infinite rapidity are nonzero.)
Furthermore Cherednikov and Stefanis do not analyze how their
definition is to be used in a valid factorization formula.

%-------------------------------------------
\subsection{Definitions in the soft-collinear effective theory (SCET) framework}

In work in SCET, a number of authors have tried to make definitions of
TMD functions.  There are two approaches.

One is that of Becher and Neubert.\cite{Becher:2010tm} They use
Smirnov's subtraction methods\cite{Smirnov:2002pj} applied in a style
appropriate for SCET.  They are able to define the \emph{product} of
TMD parton densities used in the QCD TMD factorization formula, i.e.,
they define the product $ABS$ in Eq.\ \eqref{eq:CSS.fact}.  Although
the product is called the product of two TMD parton densities, there
are, in fact, subtractions that effectively give a soft factor.  But
Becher and Neubert are unable to define individual TMD parton
densities.  As far as I can tell, their product of TMD parton
densities is the same as with my new definition; I think this is
guaranteed by the use of Smirnov's subtraction methods.

The other school in the SCET community is represented by Mantry and
Petriello.\cite{Mantry:2009qz,Mantry:2010bi}
%  Primary application to Higgs
%  More complicated formula with more variables:
They work with what they call an ``impact parameter beam function''
(iBF).  Compared with the definitions such as
\eqref{eq:TMD.pdf.pm.def} and \eqref{eq:TMD.pdf.unsub}, the fields in
the definition of an iBF are not restricted to a null plane.  So the
iBFs are actually more general objects than TMD parton densities, and
in some other literature would be called fully unintegrated parton
densities.  There are processes more exclusive than the Drell-Yan
process for which their use is advantageous or even necessary.  The
iBFs have zero bin subtractions, which are something like the soft
factors in my definition \eqref{eq:TMD.pdf.defn}.  However, the exact
relation is not clear to me.

%====================================
\section{Outlook: Implications of the new definition and associated
  derivations}

\begin{itemize}
\item We now have a precise satisfactory operator definition of TMD
  parton densities that can be taken literally.
\item The gauge links at infinity are no longer needed in Feynman
  gauge.
\item The definition is part of a new formalism with much better
  subtraction methods\cite{Collins:2011qcdbook} than were previously
  available.
\item There is now a full proof\cite{Collins:2011qcdbook} of TMD
  factorization for the Drell-Yan process.  (CSS\cite{Collins:2004nx}
  gave up on this!)
\item Hence, there is now an fully unambiguous method for calculation
  of the hard scattering coefficient.
\item The formalism is much cleaner mathematically than previous
  versions. 
\item Although I did not explain this, there is a fully specified
  relation between exact and approximated parton kinematics in the
  hard scattering.
\item As for future work, there is an urgent need to numerically
  relate TMD parton densities with the new definition to those in
  other formalisms.
\end{itemize}

%====================================
\section*{Acknowledgments}

This work was supported by the U.S. D.O.E. under grant number
DE-FG02-90ER-40577.  I would like to thank Ted Rogers for many useful
discussions.

%====================================
\bibliographystyle{ws}
\bibliography{jcc}

\providecommand{\noopsort}[1]{}
\begin{thebibliography}{10}

\bibitem{Collins:2011qcdbook}
J.~C. Collins.
\newblock {\em Foundations of Perturbative QCD} (Cambridge University Press,
  Cambridge, 2011).

\bibitem{Aybat:2011zv}
S.~M. Aybat and T.~C. Rogers, {\em Phys.Rev.}, D83:114042, 2011.

\bibitem{Cardy:1974vq}
J.~L. Cardy and G.~A. Winbow, {\em Phys. Lett.}, B52:95, 1974.

\bibitem{Collins:2004nx}
J.~C. Collins and A.~Metz, {\em Phys. Rev. Lett.}, 93:252001, 2004.

\bibitem{Collins:2003fm}
J.~C. Collins, {\em Acta Phys. Polon.}, B34:3103--3120, 2003.

\bibitem{Bacchetta:2008xw}
A.~Bacchetta, D.~Boer, M.~Diehl, and P.~J. Mulders, {\em JHEP}, 08:023, 2008.

\bibitem{Collins:1984kg}
J.~C. Collins, D.~E. Soper, and G.~Sterman, {\em Nucl. Phys.}, B250:199--224,
  1985.

\bibitem{Collins:1981uw}
J.~C. Collins and D.~E. Soper, {\em Nucl. Phys.}, B194:445--492, 1982.

\bibitem{Ji:2004wu}
X.-D. Ji, J.-P. Ma, and F.~Yuan, {\em Phys. Rev.}, D71:034005, 2005.

\bibitem{Cherednikov:2007tw}
I.~Cherednikov and N.~Stefanis, {\em Phys.Rev.}, D77:094001, 2008.

\bibitem{Cherednikov:2008ua}
I.~Cherednikov and N.~Stefanis, {\em Nucl. Phys.}, B802:146--179, 2008.

\bibitem{Cherednikov:2009wk}
I.~O. Cherednikov and N.~G. Stefanis, {\em Phys.Rev.}, D80:054008, 2009.

\bibitem{Mandelstam:1982cb}
S.~Mandelstam, {\em Nucl.Phys.}, B213:149--168, 1983.

\bibitem{Leibbrandt:1983pj}
G.~Leibbrandt, {\em Phys.Rev.}, D29:1699, 1984.

\bibitem{Becher:2010tm}
T.~Becher and M.~Neubert, {\em Eur.Phys.J.}, C71:1665, 2011.

\bibitem{Smirnov:2002pj}
V.~A. Smirnov, {\em Springer Tracts Mod.Phys.}, 177:1--262, 2002.

\bibitem{Mantry:2009qz}
S.~Mantry and F.~Petriello, {\em Phys.Rev.}, D81:093007, 2010.

\bibitem{Mantry:2010bi}
S.~Mantry and F.~Petriello, arXiv:1011.0757.

\end{thebibliography}

\end{document}